# Ultra-low threshold, electrically pumped quantum dot photonic crystal nanocavity laser

Bryan Ellis[1]*, Marie A. Mayer[2], Gary Shambat[1], Tomas Sarmiento[1], James Harris[1], Eugene E. Haller[2], and Jelena Vuckovic[1]

[1] Department of Electrical Engineering, Stanford University, Stanford, California 94305 USA
[2] Materials Sciences Division, Lawrence Berkeley National Laboratory, Berkeley California 94720 USA and Department of Materials Science, University of California, Berkeley, Berkeley, California 94720 USA

*e-mail: bryane@stanford.edu

**Efficient, low threshold, and compact semiconductor laser sources are being investigated for many applications in high-speed communications, information processing, and optical interconnects. The best edge-emitting and vertical cavity surface-emitting lasers (VCSELs) have thresholds on the order of 100 µA[1,2] but dissipate too much power to be practical for many applications, particularly optical interconnects[3]. Optically pumped photonic crystal (PC) nanocavity lasers represent the state of the art in low-threshold lasers[4,5]; however, in order to be practical, techniques to electrically pump these structures must be developed. Here we demonstrate a quantum dot photonic crystal nanocavity laser in gallium arsenide pumped by a lateral p-i-n junction formed by ion implantation. Continuous wave lasing is observed at temperatures up to 150 K. Thresholds of only 181 nA at 50 K and 287 nA at 150 K are observed - the lowest thresholds ever observed in any type of electrically pumped laser.**

PC nanocavities are an ideal platform for low-power laser sources because strong light-matter interaction can be obtained. High quality factors (>1x10$^6$) have been demonstrated in cavities with mode volumes comparable to a cubic-optical wavelength[6,7]. In such cavities the Purcell factor can be quite high, reducing the threshold and increasing the modulation rate[8]. Optically pumped PC nanocavity lasers have been demonstrated to have thresholds of only a few nW[9] and modulation rates exceeding 100 GHz[10]. In addition, they can operate in continuous wave mode at room temperature[11] and can be efficiently coupled to low-loss waveguides for optoelectronic integrated circuit applications[12]. Recently a buried heterostructure optically pumped photonic crystal laser was demonstrated in a 5.5 GHz optical communications link where the power dissipation of the laser was measured to be only 13 fJ/bit[13], which is a practical value for optical interconnects[3].

The principal disadvantage of PC nanocavity lasers is that they are intrinsically difficult to electrically pump because it is challenging to efficiently inject carriers through the membrane to the active region. For this reason, all of the aforementioned demonstrations relied on impractical optical pumping. Lasing has been demonstrated in a PC nanocavity by directing the current to the cavity region using a vertical p-i-n junction and a current post[14,15]. However, the current post limits the quality factor of the cavity, restricts the choice of the cavity design, and requires a complicated fabrication procedure[15]. In addition, a high threshold current of 260 µA was observed at a threshold voltage of around 1 V meaning that the threshold power dissipation of the laser is around 260 µW, significantly higher than in optically pumped PC devices and exceeding even that of VCSELs[2].

A lateral p-i-n junction (defined in plane of the photonic crystal) offers more flexibility than a vertical junction because the current flow can be defined lithographically to efficiently flow into the cavity region[16]. Additionally, the electrical design of the device no longer affects the optical properties. In this work we demonstrate a PC nanocavity laser that is electrically pumped by a lateral p-i-n junction as shown schematically in Figure 1a. The intrinsic region is designed to be 400 nm wide in the cavity region and extends to 5 µm wide to the sides of the cavity. This design directs the current flow through the cavity region. We choose a modified 3-hole defect PC cavity design [17]. Figure 1b shows a finite difference time domain (FDTD) simulation of the fundamental cavity mode, with theoretically estimated quality factor of 115,000, comparable to previous studies of similar cavities[18].

Ion implantation of Be and Si ions is used to dope the p- and n-type regions respectively. Because implantation of high energy ions causes some lattice damage that will reduce the gain, it is critical that the p and n regions are precisely aligned to the PC cavity to avoid damaging the active region. We developed a fabrication procedure where the ions are implanted with silicon nitride masks patterned by electron beam lithography to achieve approximately 30 nm alignment accuracy (see methods). The gain material for the laser is three layers of high density (300 dots/µm$^2$) InAs quantum dots. In order to activate the ion implanted dopants, we perform a high temperature anneal. We optimized this anneal procedure (see methods and supplementary information) to give sufficient dopant activation without significantly changing the emission properties of the quantum dots. After the activation anneal we find that the peak photoluminescence wavelength of the quantum dots is 1175 nm at 100 K. We find that the average doping density in the membrane (after dopant activation) is $6.0 \times 10^{17}$ cm$^{-3}$ and $2.5 \times 10^{19}$ cm$^{-3}$ in the n and p regions respectively (see supplementary information).

To confirm the doping layout before fabrication of the PC cavity we use scanning capacitance atomic force microscopy (SCM). Figures 2a and 2b show the SCM topography and SCM data for structures without PCs demonstrating that the desired dopant layout is achieved. From this data we can identify the exact locations of the doping regions and precisely position the PC cavity in the center of the intrinsic region (see supplementary information). Figures 2c and d show SEM images of the fabricated PC laser. The parameters of the cavity are chosen so that the fundamental cavity mode is at a wavelength of 1174 nm at low temperature, within the ground state emission of the quantum dots. We confirm that this mode is the fundamental mode of the cavity by identifying the higher order cavity modes in the electroluminescence spectra and comparing them with FDTD results. Figure 2e shows the current-voltage characteristics of the fabricated laser diode at a temperature of 50 K.

Figure 3a shows the optical output power of the laser as a function of continuous wave pump current at several different temperatures. We observe a clear lasing threshold for temperatures below 150 K. As the temperature increases, the quantum dot resonance wavelength and the cavity resonance wavelength red-shift at different rates. At higher temperatures, the photonic crystal cavity resonances are no longer within the gain bandwidth of the quantum dots. A measurement on the same structure at 200 K where no lasing is observed is included for comparison. To determine the threshold of our laser we use a linear fit to the above threshold characteristics and extrapolate it to zero-output power (red lines in figure 3a). We find that the threshold of our laser is 181 nA at 50 K and 287 nA at 150 K. To the best of our knowledge, this is the lowest threshold ever demonstrated in an electrically pumped semiconductor laser. It is 3 orders of magnitude better than the 260 µA threshold demonstrated in quantum well PC cavity lasers[14] and more than an order of magnitude better than the thresholds demonstrated in metal-clad lasers[19] and micropost lasers[20]. At threshold, the applied voltage is only 1.15 V at 50 K and 1.03 V at 150 K, meaning that at threshold the lasers consume only 208 nW at 50 K and 296 nW at 150 K. We estimate the total power radiated by the laser to be on the order of tens of pW well above threshold[21]. Figure 3b shows the experimental far-field radiation pattern of the cavity at various current levels.

The laser linewidth (full width half maximum) as a function of current is plotted in Figure 4a. The linewidth narrows from around 1.35 nm just below threshold to 0.95 nm well above threshold. The quality factor of the cavity mode at threshold is approximately 1130. This quality factor is significantly smaller than expected from FDTD simulations and may be too low to achieve lasing at room temperature. The low quality factor is likely due to surface roughness introduced by the fabrication procedure that increases the out-of-plane radiation loss of the cavity (see supplementary information). Room-temperature optically pumped lasing in nanocavities with similar quantum dots has recently been demonstrated[11,22], and we believe that further refinements to the fabrication procedure will allow us to demonstrate electrically pumped lasing at room temperature.

Although we have optimized our device design to reduce leakage current (see methods), at low voltages before the diode has fully turned on we observe leakage current bypassing the cavity through the sacrificial layer and substrate. Therefore, if the device design is further improved to reduce this leakage, the threshold could be significantly lower. To find the potential threshold reduction, we fit the current voltage characteristics to an ideal diode equation to determine the fraction of current flowing through the cavity as a function of applied voltage (see supplementary information). The light output of the laser as a function of the current after subtracting the leakage is plotted in Figure 4b (blue points) along with a fit to the laser rate equations (see supplementary information). From the fit we determine that the fraction of spontaneous emission that is coupled to the cavity mode (commonly called the β-factor[10]) is approximately 0.61 in our laser. The laser threshold after correcting for the leakage current is only 70 nA.

In summary, we have designed and demonstrated an electrically pumped quantum dot photonic crystal nanocavity laser. The laser operates in continuous wave mode at temperatures up to 150 K, and exhibits ultra-low thresholds of 181 nA at 50 K and 287 nA at 150 K. If we subtract the leakage current flowing into the substrate, the threshold current is estimated to be around 70 nA. These lasing thresholds are three orders of magnitude lower than previous demonstrations of electrically pumped PC nanocavity lasers, and lower than any electrically injected laser so far. We believe that room temperature operation is possible if the quality factors of the cavity can be improved, and if the cavity resonances are better aligned to the quantum dot gain spectrum at room temperature. The low power dissipation of these lasers makes them very promising for applications in optical interconnects and high speed communications as well as for fundamental studies of the properties of electrically pumped thresholdless lasers and lasers with single-emitter gain[23].

**Acknowledgements**

Bryan Ellis and Gary Shambat were supported by the Stanford Graduate Fellowship.  Gary Shambat is also supported by the NSF GRFP.  The authors acknowledge the support of the Interconnect Focus Center, one of six research centers funded under the Focus Center Research Program (FCRP), a Semiconductor Research Corporation entity.  The authors would like to acknowledge Ilya Fushman for helpful discussions and Mark Hilton of Veeco Instruments for advice regarding SCM.  Work was performed in part at the Stanford Nanofabrication Facility of NNIN supported by the National Science Foundation.

**Author Contributions**

B.E. and J.V. designed the experiment. T.S. and J.H. performed the MBE growth of the samples. B.E. and M.M. fabricated the devices. B.E., M.M., and E.H. characterized the fabricated samples. B.E. and G.S. performed the measurements. B.E. analyzed and modeled the data. B.E. and J.V. wrote the paper. All authors contributed to discussions.

**Competing Financial Interests Statement**

The authors declare no competing financial interests.

**Methods**

**Wafer Growth** – The wafer was grown using molecular beam epitaxy. Starting with a semi-insulating substrate, a 1 µm $Al_{0.95}Ga_{0.05}As$ sacrificial layer was grown, followed by a 220 nm GaAs membrane that contained three layers of InAs quantum dots separated by 50 nm GaAs spacers. The dots were formed by depositing 2.8 monolayers of InAs at 510 $^{\circ}$C using a growth rate of 0.05 monolayers/s. The dots were capped with a 6 nm $In_{0.15}Ga_{0.85}As$ strain reducing layer. The resulting dot density was approximately 300 dots/µm$^2$ as confirmed by atomic force microscopy measurements of uncapped quantum dot samples.

**Fabrication** – First, alignment marks were defined on the unpatterned wafer using electron beam lithography and dry-etched around 100 nm into the membrane using an $Ar/Cl_2/BCl_3$ electron-cyclotron resonance reactive ion etch (ECR-RIE). Next, a 330 nm layer of silicon nitride was deposited on the sample using plasma-enhanced chemical vapor deposition (PECVD) to serve as a mask for ion implantation of Si.  Electron beam lithography was used to pattern the n-type doping region, and an SF$_6$/C$_2$F$_6$ dry etch was used to remove the nitride from the n-type doping area.  Si ions were implanted at an energy of 115 keV and a dose of 3e14/cm$^2$.  An SF$_6$/C$_2$F$_6$ dry etch was used to remove the remaining silicon nitride, and another 330 nm layer of silicon nitride was deposited on the sample using PECVD to serve as the mask for ion implantation of Be.  Electron beam lithography was used to pattern the p-type doping region and an SF$_6$/C$_2$F$_6$ dry etch was used to remove the silicon nitride from the p-type doping area.  Be ions were implanted at an energy of 32 keV and a dose of 2.5e15/cm$^2$.  An SF$_6$/C$_2$F$_6$ dry etch was used to remove the remaining silicon nitride.  A 40 nm tensile strained silicon nitride cap was deposited using PECVD to prevent As out-diffusion during the subsequent high temperature anneal.  The samples were then annealed at 850$^\circ$C for 15 s in a rapid thermal annealer to activate the dopants and remove almost all of the lattice damage caused by the ion implantation.  An SF$_6$/C$_2$F$_6$ dry etch was used to remove the nitride cap.  The photonic crystal pattern was defined using electron beam lithography and etched into the membrane using an Ar/Cl$_2$/BCl$_3$ ECR-RIE.  Simultaneously with the photonic crystal, trenches were etched to the sides of the cavity and all the way around each of the contacts; this was found to reduce the leakage current to reasonable levels.  Next, the photonic crystal was loaded in a wet thermal oxidation furnace and the sacrificial layer was oxidized at 465 $^\circ$C for 7 minutes.  Photolithography and electron beam evaporation were used to define Au/Ge/Ni/Au n-type contacts in a lift-off process.  Photolithography and sputtering were used to define Au/Zn/Au p-type contacts also in a lift-off process.  The contacts were then annealed at 415 $^\circ$C for 15 s to achieve minimum contact resistance.  Finally, 45% potassium hydroxide solution in water was used to remove the oxidized sacrificial layer underneath the cavity leaving an air-clad photonic crystal membrane.

**Sample Characterization –** Hall effect was used to find the carrier sheet concentration in semi-insulating GaAs test samples implanted with the same conditions as the laser samples.  The doping profile was measured using an electrochemical capacitance voltage measurement using 0.1 M NaOH with EDTA surfactant as the electrolyte.  An atomic force microscope with a scanning capacitance attachment was used to measure the doping layout on the actual laser samples on devices without photonic crystals.  A two plate capacitor setup for scanning consisted of the sample, native surface oxide, and a gold coated AFM tip used in contact mode with a scanning bias of 1 V.

**Sample Optical Testing –** The sample was epoxied to an alumina chip carrier using nonconductive, vacuum-safe epoxy.  Aluminum wirebonds were used to connect individual devices to the leads of the chip carrier, and the chip carrier was loaded into a continuous flow helium cryostat with custom designed coldfinger and electrical feedthroughs.  The temperature was stabilized to within half a degree Kelvin.  Currents were applied using a sourcemeter with sub nA accuracy.  The emission from the sample was collected using an objective lens with numerical aperture 0.5 in the direction perpendicular to the sample surface.  Emission spectra were measured using a liquid-nitrogen cooled spectrometer with InGaAs charge coupled device (CCD) detector, and luminescence images were taken with an InGaAs CCD camera.

**Supplementary Information**

**I.  Laser material characterization**

As mentioned in the main text, a high temperature anneal is necessary to activate the ion implanted dopants and remove most of the lattice damage caused during implantation. In order to optimize the annealing conditions to achieve a high doping density, we implanted semi-insulating GaAs samples under the same conditions as the laser samples (implantation conditions are described in the methods section). We tested the activation efficiency of our dopants for rapid thermal anneal temperatures between 800 $^{o}$C and 900 $^{o}$C for a range of times between 10 and 30 s. We found that Be activates most efficiently at 800 $^{o}$C while Si activates most efficiently at 900 $^{o}$C. We chose an intermediate anneal temperature of 850 $^{o}$C for 15 s. Room temperature Hall effect measurements were used to characterize the doping density of the test samples. Using Hall effect, we found a sheet carrier density of 7.9*10$^{14}$ cm$^{-2}$ (1.1*10$^{13}$ cm$^{-2}$) and a mobility of 126 cm$^2$/Vs (1930 cm$^2$/Vs) for the p-type (n-type) sample. The high mobility values indicate that the anneal step was effective in removing much of the lattice damage. Electrochemical capacitance voltage (ECV) measurements were used to determine the doping density as a function of depth. Supplementary figure SF1 shows the doping density measured on both the p-type and n-type test samples. The sheet carrier density found by integrating the ECV data is comparable to the sheet density measured using Hall effect.

    The high temperature anneal also affects the emission properties of the quantum dots. Previous studies of this effect have found that high temperature anneals tend to blueshift the quantum dot photoluminescence and narrow the inhomogeneous broadening of the quantum dot ensemble [S1,S2]. This could be beneficial for laser applications because theoretically the quantum dot gain could be increased by annealing. The effects of the activation anneal on the quantum dots are summarized in Supplementary figure SF2. Figure SF2 shows the normalized

quantum dot photoluminescence at 100K for three different annealing conditions.  The blue line shows the photoluminescence of the sample without anneal, while the red and black lines show the photoluminescence after annealing at 850 °C for 15 s and 30 s respectively.  We observe no significant narrowing of the photoluminescence full width half maximum, but there is a significant blueshift.  The photonic crystal cavities are designed to have the fundamental mode resonant with the ground state of the quantum dot photoluminescence after annealing and at low temperature.

It has been found that the diffusion coefficient of Be in GaAs is much higher than that of Si, especially for Be concentrations above $10^{19}$ cm$^{-3}$ [S3].  Therefore, it was expected that the activation anneal will redistribute the dopants significantly.  We used high resolution scanning capacitance microscopy (SCM) to approximate the extent of dopant diffusion.  The technique relies on the fact that the silicon nitride implant mask introduces some surface roughness that is visible on the topography scan of the SCM data.  This surface roughness can then be compared to the location of the edge of the doping to determine the amount of dopant diffusion during the activation anneal.  Supplementary figure SF3 presents the results of a high resolution SCM scan taken at the center of the data presented in Figure 2a and Figure 2b.  Figure SF3a shows the topography image where the black dashed lines indicate the location of the edge of the nitride mask.  The magnitude of this surface disruption is less than 1nm.  The deep pits visible in the topography image have been identified as surface defects originating during the MBE growth of the material.  Figure SF3b shows the corresponding SCM data.  The black dashed lines indicate the edge of the p and n-type doping regions.  By comparing the location of the doping region edges with the edges of the nitride mask, we determine that the Be doping edge is approximately 300 nm from the mask edge and the Si doping edge is approximately 150 nm from the mask edge. This allows us to determine the precise location of the dopants and position the photonic crystal cavity directly in the center of the doping.

Figure SF3c shows a linescan of the topography data taken at the white dotted line of Figure SF3a. The linescan shows that the surface roughness is approximately 1 nm. AFM scans on the surface of the material after MBE growth indicate that the surface roughness before fabrication is approximately 0.22nm. Likely causes of this roughness include strain between the cap layer and the GaAs during the high temperature anneal and the many dry-etching steps involved in the fabrication of the device. We believe that this surface roughness is limiting the quality factor of our photonic crystal cavities. Thus, improvements in the fabrication procedure should improve the quality factor of the cavities .

## II. Electrical characterization of the photonic crystal laser

The fabricated laser diodes have some leakage current likely due to current bypassing the active region of the laser through the substrate. This leakage current is visible as a deviation from the behavior of an ideal diode at low voltages. Supplementary Figure SF4a shows the current-voltage (I-V) curve of the laser diode at 50 K plotted on logarithmic scale. At a voltage of approximately 1.2 V, the majority of the current is passing through the active region of the device, and the IV curve becomes like that of an ideal diode. Equation (1) describes the current flowing through an ideal diode with a series resistance $R_s$ and ideality factor n.

$$I = I_o \left( exp\left(\frac{q(V-IR_s)}{nk_bT}\right) - 1 \right) \tag{1}$$

The red plot on figure SF4a is a fit of the IV curve to equation (1). I in equation (1) represents the amount of current that is flowing through the active region of the diode as a function of the applied voltage. This technique is not completely accurate especially at low current values, because at threshold one would expect a small discontinuity in the IV curve as the laser transitions from spontaneous emission to stimulated emission [S4]; however we can confirm that this fitting method is reasonably accurate by plotting the integrated spontaneous emission intensity as a function of the current flowing through the cavity (after subtracting the laser mode intensity), and this is plotted in figure SF4b. Below threshold the spontaneous emission is a linear function of the current flowing through the cavity. Above threshold the spontaneous emission is slightly sub-linear because the carrier density interacting with the laser mode is clamped above threshold. From the fit we obtain $I_o=1.36*10^{-11}$ nA, $R_s=1.153$ k$\Omega$, and n=8.89. This corrected current is used to plot the light out of the laser as a function of the current flowing through the cavity in figure 4.

One would expect that for a p-i-n junction with surface recombination the ideality factor of the diode should be 2. We observe that the ideality factor decreases as temperature increases. We measure an ideality factor of 2.69 at room temperature. High ideality factors have been observed in GaN p-n junctions and have been attributed to rectifying metal-semiconductor junctions [S5]. We suspect that the metal contacts are no longer ohmic at low temperatures due to the low doping density on the n-side of the device causing an increase in ideality factor; however more studies are necessary to confirm this.

### III. Fit to the laser rate equations

In order to extract the properties of the laser we fit the optical output of the laser as a function of the current after correcting for the leakage to the laser rate equations [S4]. The laser rate equations are given by equations (2) and (3).

$$\frac{dN}{dt} = \frac{\eta I}{qV} - \frac{N}{\tau_{sp}} - \frac{N}{\tau_{nr}} - GP \qquad (2)$$

$$\frac{dP}{dt} = \Gamma GP + \frac{\Gamma \beta N}{\tau_{sp}} - \frac{P}{\tau_p} \qquad (3)$$

where η is the fraction of current injected into the active region, I is the pump current, V is the active volume, $\tau_{sp}$ and $\tau_{nr}$ are the spontaneous and nonradiative recombination rates, β is the fraction of spontaneous emission that is coupled to the laser mode, Γ is the confinement factor, P is the photon density and N is the carrier density. For the gain we use a logarithmic gain model given by equation (4).

$$G = G_o \ln\left(\frac{N}{N_{tr}}\right), \qquad (4)$$

where $G_o$ is the gain coefficient and $N_{tr}$ is the transparency carrier density. The spontaneous emission lifetime of the employed quantum dots is estimated from the literature to be approximately 3ns [S6]. The nonradiative lifetime is too long to significantly affect the fit. The active volume is $3 \ast 10^{-14}$ cm$^3$, and the photon lifetime is estimated from the quality factor of the cavity at threshold to be 0.62 ps. Γ is estimated from the overlap between the mode volume and the active volume to be 0.01. The rate equations are fit with β,η,$G_o$, and $N_{tr}$ as variable parameters. From the fit we obtain β=0.61, η=0.0069, $G_o$=1.48$\ast 10^{17}$ s$^{-1}$, and $N_{tr}$=4.12$\ast 10^{14}$ cm$^{-3}$.

**Figure Legends**

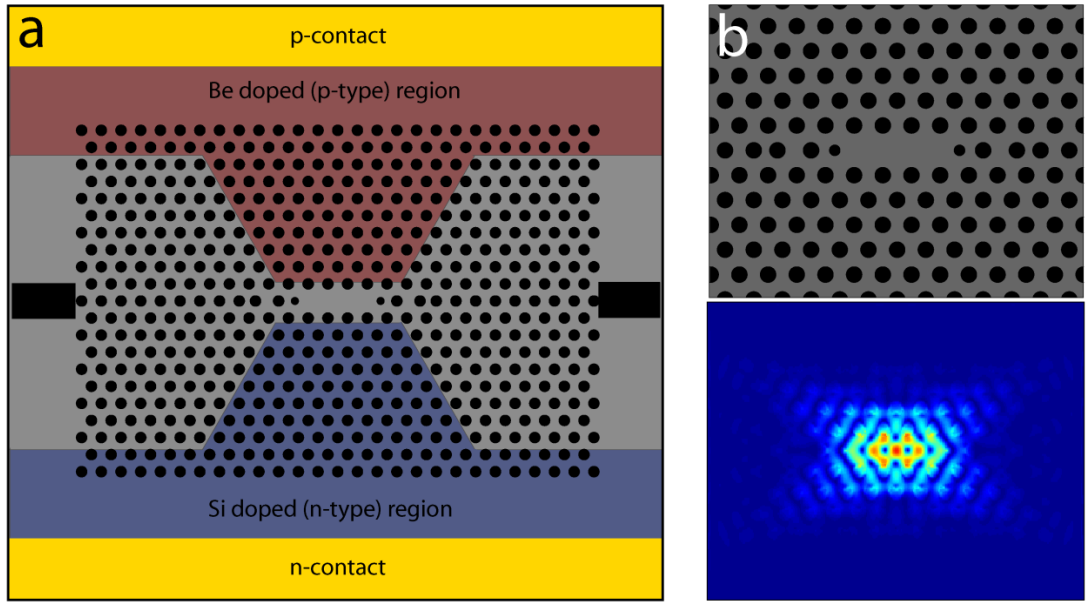

**Figure 1 | Design of the electrically pumped photonic crystal laser. a**, Schematic diagram of the electrically pumped photonic crystal laser. The p-type (n-type) doping region is indicated in red (blue). The intrinsic region width is narrow in the cavity region to direct current flow to the active region of the laser. A trench is added to the sides of the cavity to reduce leakage current (see methods). **b**, The modified three hole defect photonic crystal cavity design (top) and a finite difference time domain simulation of the E-field of the cavity mode in such a structure (bottom).

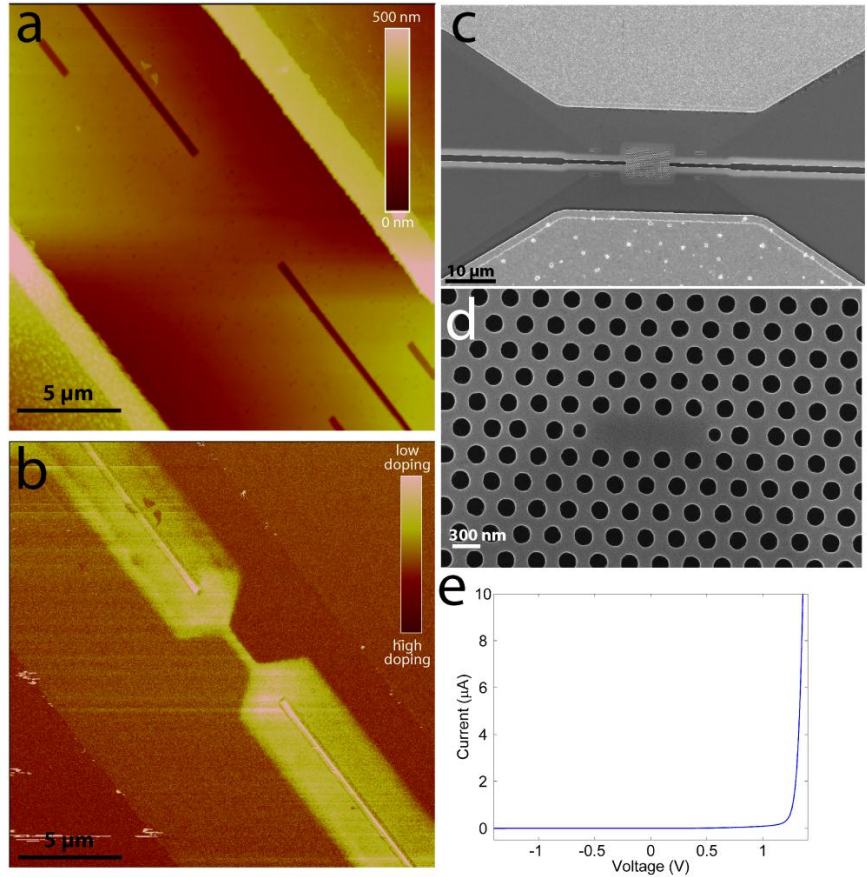

**Figure 2 | Fabrication and characterization of the photonic crystal laser device**. **a**, SCM topography image of fabricated device without photonic crystal. **b,** SCM image of the same device as in part a. The p-side (n-side) of the device is in the lower left (upper right) corner. The trench is etched at device center, showing the precision of the alignment of the doping regions. **c,** Scanning electron microscope (SEM) image of the fully fabricated laser. The p-side (n-side) of the device appears on the top (bottom) of the image. **d,** SEM image of the photonic crystal cavity (zoom-in of the central region of Fig. 2c). **e,** Current-voltage characteristics of the laser taken at 50K in the dark.

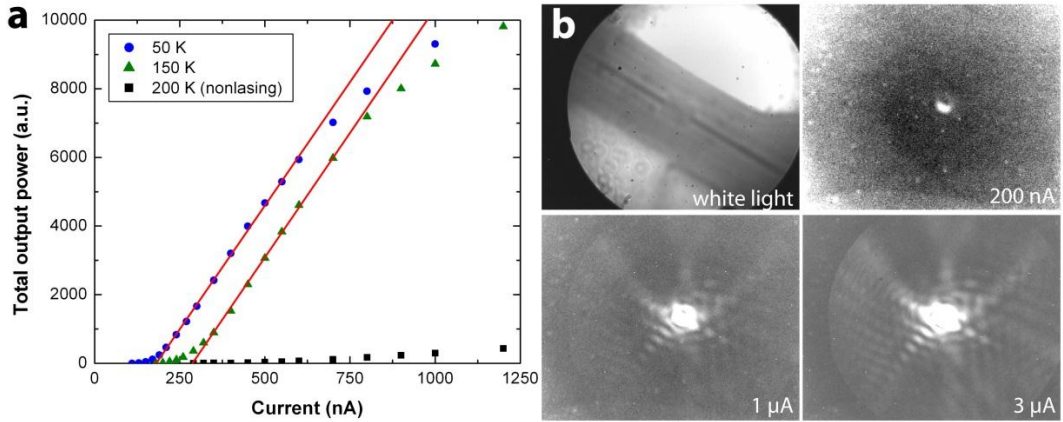

**Figure 3 | Optical output of the photonic crystal laser. a,** Experimental output light power as a function of the current through the laser at 50 K (blue points), 150 K (green points), and 200 K (nonlasing - black points). The red lines are linear fits to the above threshold output power of the lasers, which are used to find the thresholds. **b,** Far field radiation patterns of the laser at currents of 200 nA, 1 µA, and 3 µA taken at 50 K. A white light image of the cavity is also shown. The n-contact (p-contact) is in the lower left (upper right) corner of the image.

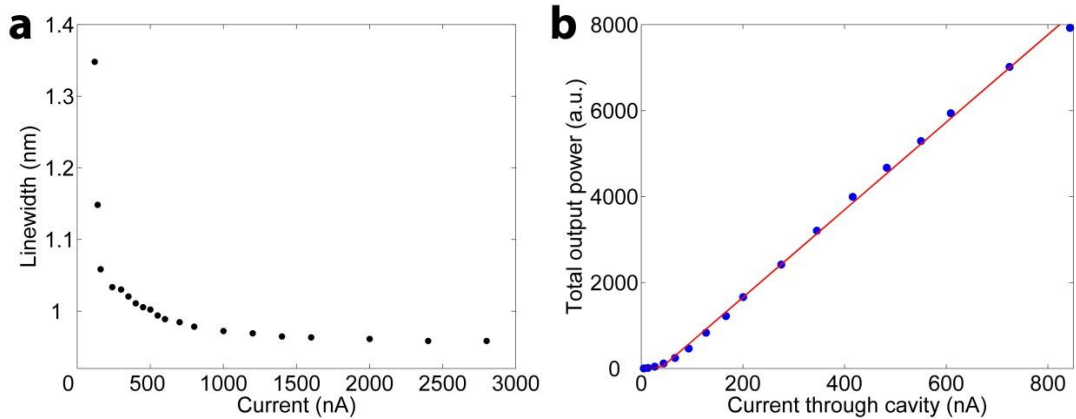

**Figure 4 | Optical properties of the photonic crystal laser. a,** Linewidth of the photonic crystal laser as a function of the current. **b,** This figure re-plots the data shown in Figure 3a for 50 K after subtracting the leakage current of the diode (blue points); the result is then fitted to the

laser rate equations (red line). "Current through cavity" shown on the horizontal axis refers to the current flowing through the cavity region after subtracting leakage through the substrate (see text and supplementary information).

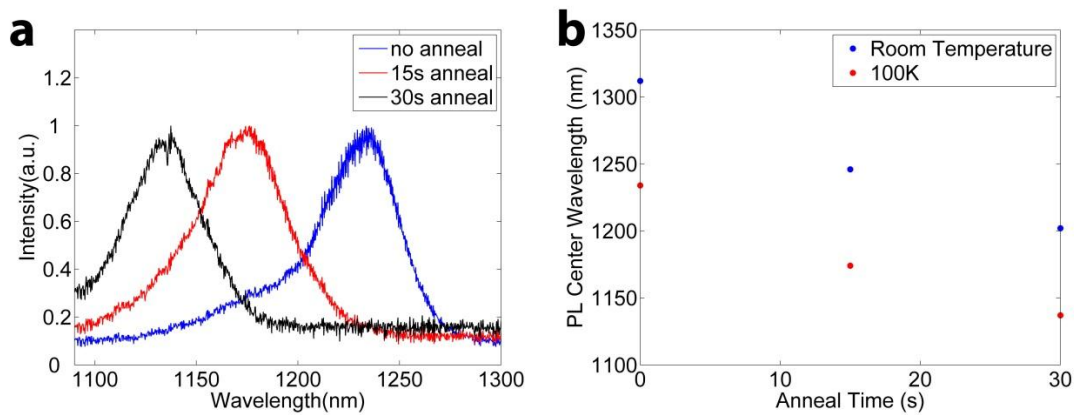

**Supplementary Figure 1 | Optical properties of InAs quantum dots after the high-temperature anneal. a,** Normalized photoluminescence of the InAs quantum dot ensemble after annealing at 850 °C for 0 s (i.e., no anneal - blue), 15 s (red), and 30 s (black). **b,** Center wavelength of the photoluminescence peak as a function of anneal time measured at both room temperature and 100 K.

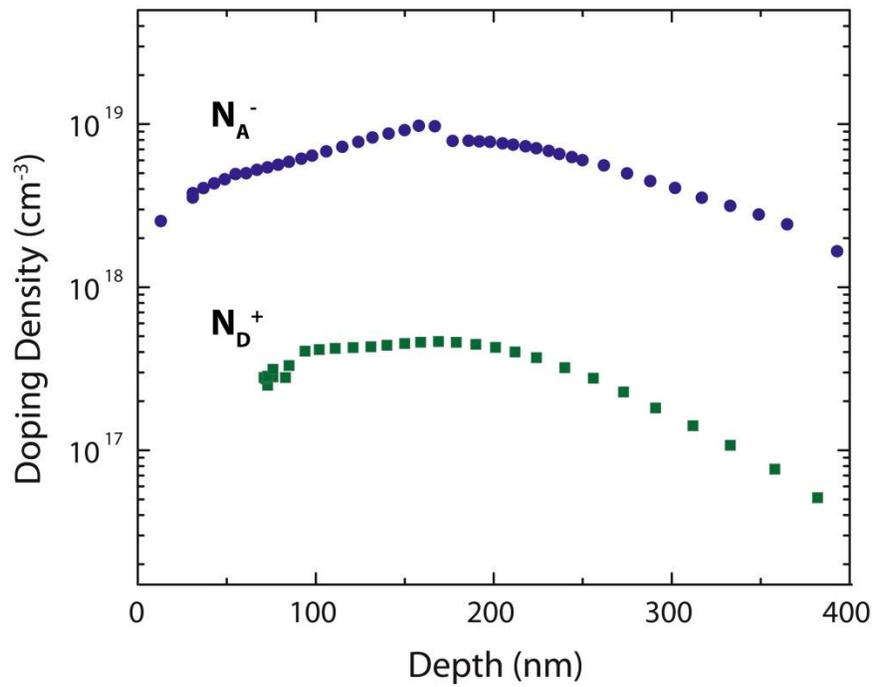

**Supplementary Figure 2 | Electrochemical capacitance voltage measurement of doping concentration as a function of depth.** The p-type sample is shown with green points and the n-type sample is shown with blue points.

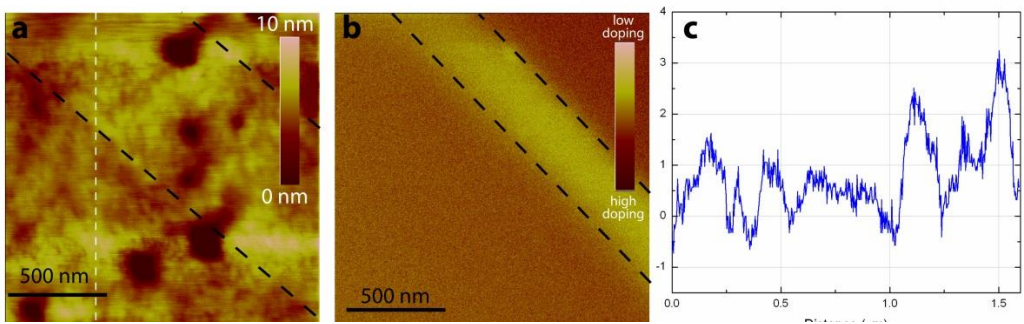

**Supplementary Figure 3 | Characterization of the diffusion of ion implanted dopants during the high-temperature anneal. a,** SCM topography image of the center region shown in Figures 2a and 2b. The edge of the nitride implantation masks are visible as faint lines. Black dashed lines are added at the location of the mask edges as a guide to the eye. **b,** SCM data showing

the location of the edges of the doping.  Black dashed lines are added as a guide to the eye to indicate the doping edges.  **c,** Linescan of the topography at the location of the white dashed line in part **a** showing >1 nm surface roughness.

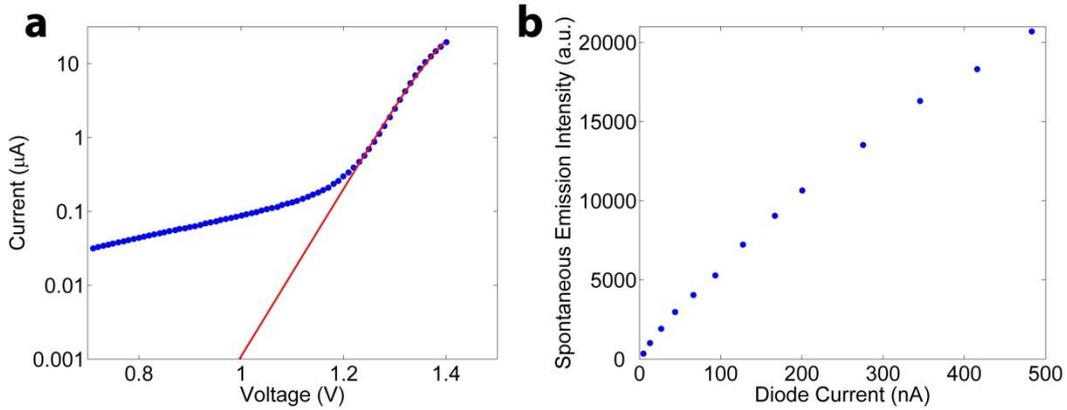

**Supplementary Figure 4 | Fit to the ideal diode curve to determine the leakage current.  a,** Current as a function of the applied voltage of the laser taken at 50 K (blue).  At high voltages (>1.15 V) the current is dominantly flowing through the cavity region.  At low voltages (< 1.15 V) most of the current is due to leakage current.  The red line is a fit to the ideal diode equation (equation (1)).  **b,** Integrated spontaneous emission intensity (after subtracting the laser intensity) as a function of the current after correcting for the leakage current.